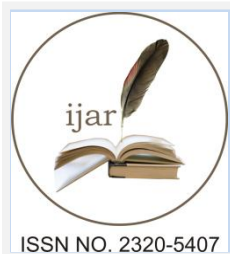
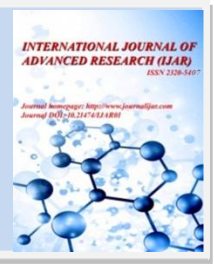

*RESEARCH ARTICLE*

## EVALUATION OF THE MANAGEMENT OF HOSPITAL RADIOLOGICAL PROTECTION

**Andrea Huhn[1], Mara Ambrosina Oliveira Vargas[2], Jorge Lorenzetti[3], Franciele Budziareck das Neves[4], Patrícia Fernanda Dorow[5], Laurete Medeiros Borges[6], Luís Lança[7], Carlos Queiroz[8].**

1. Doctoral student PEN/UFSC. Professor of the Undergraduate Program in Radiology Technology and the Master Degree in Radiological Protection of the Federal Institute of Santa Catarina (IFSC). Florianópolis, Santa Catarina, Brazil.
2. Ph.D. in Nursing Philosophy. Professor, Nursing Department and PEN/UFSC. Florianópolis, Santa Catarina, Brazil.
3. Ph.D. in Nursing. Professor, Nursing Department and PEN/UFSC. Professor, Nursing Department and PEN/UFSC. Florianópolis, Santa Catarina, Brazil.
4. Doctoral student at the Graduate Program in Nursing (PEN), Federal University of Santa Catarina (UFSC). Florianópolis, Santa Catarina, Brazil.
5. Ph.D. in Engineering and Knowledge Management. Professor of the Undergraduate Program in Radiology Technology and the Master Degree in Radiological Protection of the Federal Institute of Santa Catarina (IFSC). Florianópolis, Santa Catarina, Brazil.
6. Ph.D. in Nursing. Professor, Nursing Department and PEN/UFSC. Professor of the Undergraduate Program in Radiology Technology and the Master Degree in Radiological Protection of the Federal Institute of Santa Catarina (IFSC). Florianópolis, Santa Catarina, Brazil.
7. Radiographer. Phd Helth Techolonogies (ESTeSL).
8. System programmer

……………………………………………………………………………………………………....

*Abstract*
………………………………………………………………
To maintain quality in hospital services, management strategies are fundamental. The objective of this article was to elaborate and validate the contents of an instrument for the management of hospital radiological protection. Therefore, a study was conducted in two Portuguese-speaking countries, Brazil and Portugal. Initially, a data collection instrument was created to elaborate essential items for the management of hospital radiological protection, followed by the validation of the contents of this instrument, using the Delphi technique. The validation of the instrument content was performed by a group of judges, following the steps of the Delphi technique. The questionnaire answered 33 professionals, of these, 25 Brazilians and 8 Portuguese. The affirmative statements among the professionals are related to the instructions on radiological protection for the radiodiagnostic team and the radiological protection program. It is concluded that the instrument built and validated for the management of radiological protection contributes to the organization of diagnostic imaging services and may be adapted for the management of specific services.

……………………………………………………………………………………………………....

**Corresponding Author:-Andrea Huhn.**
Address**:-**Doctoral student PEN/UFSC. Professor of the Undergraduate Program in Radiology Technology and the Master Degree in Radiological Protection of the Federal Institute of Santa Catarina (IFSC). Florianópolis, Santa Catarina, Brazil

**Introduction:-**
When reflecting on the quality of hospital services, it is essential to discuss the management strategies that support the institutions, so that they can meet the needs and demands of the worker and user in all its dimensions. Since the 1930s, there have been debates about the improvement in the quality of hospital services, since decision-making in the health sector requires an awareness of professionals and managers, since they live in a constantly changing scenario of care to ensure quality patient care and services. Since then, the creation of systems, cheklists, fiches, protocols, questionnaires, programs and policies have been gaining space and visibility in the discussions about hospital management planning (Feldman et al., 2005; Cantiello et al., 2016). Among the various sectors of a hospital there is radiodiagnosis, an extremely complex sector because it has different modalities of image acquisition that is used as one of the main tools for diagnosis of pathologies, monitoring of treatment and prediction of results, through ionizing radiation. Considering that ionizing radiation interacts with matter, ionizes its atoms or molecules, it is imperative that the multiprofessional health team working in this sector know and apply the principles of radiation protection, especially the principle of ALARA (Huhn et al., 2017).

The ALARA principle "as low as reasonably achievable", as described by the International Commission on Radiological Protection (ICRP), recommends using the lowest radiation dose possible without affecting image quality to protect both the health of the occupationally exposed individual, and of the individual (user) who needs an accurate diagnosis. In addition, the International Commission on Radiological Protection recommends that steps be taken to reduce unnecessary exposures (Icrp, 1977).

That is, an examination must follow a pre-established protocol to avoid repetition and consequently excessive dose of ionizing radiation and this question can only be reached with a management appropriate to the radiological protection.

In this regard, it is worth remembering that the primary objective of radiological protection is to provide a protection standard for individuals without limiting beneficial practices involving radiation exposure, also taking into account that radiological protection should be equally relevant for imaging tests. low or high complexity (Iaea 2004). Following the above assumptions, it is possible to build an effective Radiological Protection Program, making adequate use of the technologies that use ionizing radiation, without causing damage to the health of workers and users of the diagnostic imaging service.

In order for this to be effective in practice, it is necessary for the institutions that offer radiodiagnostic service to attend to the management of radiological protection, to provide educational and easily accessible resources related to the importance of radiological protection for multiprofessional staff and users (Huhn et al. al., 2017).
In addition, it is necessary to make available educational and updated documents and materials in this area of knowledge, as well as providing spaces for the exchange of ideas between the multiprofessional team, be they physical or virtual, through the available information technologies that are now easily accessible. These resources result in good radiation protection practices (Brand et al., 2009).

In this sense, the justification for conducting this study is based on the fact that, to date, there is no knowledge of the existence of any instrument that guides the management of hospital radiological protection. However, in other segments of the health area, there is an increasing and significant number of questionnaires and scales that aim to evaluate a study phenomenon (Coluci et al., 2015). Thus, the objective was to elaborate and validate the content of an instrument for the management of hospital radiological protection in two Portuguese-speaking countries, respectively, Brazil and Portugal.

**Methods:-**
This is a methodological study, consisting of two parts: a) construction of a data collection instrument for the preparation of essential items for the management of hospital radiological protection and the validation of the contents of this instrument, using the Delphi technique. The validity of content verifies if the instrument measures exactly what it proposes, that is, it evaluates the ability of an instrument to accurately measure the phenomenon in question (Munaretto et al., 2013, Hulley et al., 2015). The Delphi technique seeks agreement in the opinions of a group of judges on a particular topic and can be performed through paper and forms collected by physical mail or can also be performed by digital means via e-mail being called e-Delphi (Keeney et al., 2011; Massaroli et al., 2017), which is the case of this study. The validation of the content of the instrument was performed by a group of judges, made up of Brazilian and Portuguese radiology technologists, following the steps of the Delphi technique (Munaretto et al., 2013; Keeney et al., 2011).

The definition of the problem is given by the following question: How significant is PR management in the hospital radiology service? A Determination of the necessary knowledge of the participants was: To have training in the area of radiodiagnosis, specifically to be a Radiology Technologist with experience in hospital radiodiagnosis. The selection criteria of the judges were as follows: to be a radiology technologist trained at least five years ago, to have worked in hospital radiodiagnosis in Brazil or Portugal for at least one full year.

A total of 60 technologists in radiology were selected, with 30 Brazilian professionals and 30 Portuguese professionals. It should be noted that the authors of this study indicated five professionals from each country and the other participants were nominated by these professionals, successively. All the professionals indicated received a personal and / or professional email, a LimeSurvey questionnaire with information about the objectives and description of the study, as well as their rights as participants. Of the total, 33 professionals answered the questionnaire, 25 Brazilians and 8 Portuguese.

The questionnaire was available for response for 120 days and all selected professionals received reminders every 15 days reinforcing the invitation to study, as well as the objectives and Free and Informed Consent Term (TCLE). The study was approved by the Ethics and Research Committee with Human Beings under opinion No. 1938144. The entire study was guided and obeyed the ethical care set forth in Resolution 466/2012 of the National Health Council.

The construction of the data collection instrument was guided by a previously prepared checklist, based on an analysis of the legislation regarding PR and non-participant observation of the routine of a hospital radiodiagnosis service in southern Brazil. After that, the cheklist was transformed into a LimeSurvey questionnaire, in agreement between two researchers, radiology technologists (one from Brazil and one from Portugal), both with more than 15 years of training and experience in hospital diagnostic radiology. In addition, in order to evaluate the questionnaire by Brazilians and Portuguese, the principles of radiological protection adopted internationally were taken into consideration in order to evaluate the consensus of the judges regarding the management of hospital RH in Brazil and Portugal.

Fifteen statements were made questioning its relevance for the management of hospital radiation protection. The affirmatives contained 4 options of response, in scale of the type Likert, that in this study was very relevant to nothing relevant, being obligatory the answer to advance to the next question. After each statement, a space was available for comments on its relevance to radiological protection, if the judge judged it necessary (for the composition of the results, the judges' observations on the statements were identified as Affirmative 1 = A1, Affirmative 2 = A2, and so on). This field was not mandatory to proceed with the questionnaire.

Step of data collection itself. In this case, the questionnaire was sent by e-mail. Data collection took place between May and September 2017. The content validity index (CVI) and the Kappa coefficient (K) were used to analyze the relevance of the questionnaire using the Delphi technique. IVC is widely used in health to measure the proportion / percentage of judges who are in consensus on the items of an instrument and its aspects.13-14 In this study it was used to analyze each item individually and finally the instrument as a everything. A Likert-type scale with a score of one to four was used to assess the relevance of the responses3, where it was considered: 1 = Extremely relevant; 2 = Very relevant; 3 = Not relevant; 4 = Not relevant.

For this study, a IVC with an acceptable concordance rate of 0.80 was adopted, corroborating with the literature, which establishes that in the evaluation process of the individual items, the number of judges should be considered and, if there are six or more, a rate of not less than 0,78 is recommended. To verify the validity of new instruments in general, a minimum agreement of 0.80 is suggested (Wynd et al., 2003; Davies et al., 2018).

Fleiss's kappa coefficient has been very useful for evaluating measures of concordance among judges in the health area (Wynd et al., 2003). This study was calculated by a calculator available online (http://justusrandolph.net/kappa/). A number equal to or greater than 0.80 was used. For concordance analysis, the four possible ordinal responses on the Likert scale were grouped into two categories. It was considered that there was agreement between the judges who marked alternatives 1 or 2 on the Likert scale (relevant item), as well as 3 or 4 (item without relevance). (Wynd et al., 2003, Davies et al., 2018).

The consensus was obtained immediately, so no other rounds were performed, as recommended by the Delphi technique, when the consensus in the answers does not occur in the first round. In this direction, the next step was the compilation of the answers and presentation of the results, which are presented in the results section, below.

**Results:-**
Table 1 shows the results obtained by the Kappa Coefficient and the content validity index.

**Table 1:-**Results obtained by Kappa Coefficient and IVC

| Question | Extremely / Very Relevant | Little/ Anything relevant | Kappa de Fleiss | IVC |
|---|---|---|---|---|
| 1 | 31 | 2 | 0.88 | 0,93 |
| 2 | 30 | 3 | 0.83 | 0,90 |
| 3 | 32 | 1 | 0.94 | 0,96 |
| 4 | 30 | 3 | 0.83 | 0,90 |
| 5 | 32 | 1 | 0.94 | 0,96 |
| 6 | 31 | 2 | 0.88 | 0,93 |
| 7 | 31 | 2 | 0.88 | 0,93 |
| 8 | 31 | 2 | 0.88 | 0,93 |
| 9* | 28 | 5 | 0.73 | 0,84 |
| 10 | 31 | 2 | 0.88 | 0,93 |
| 11 | 32 | 1 | 0.94 | 0,96 |
| 12 | 32 | 1 | 0.94 | 0,96 |
| 13 | 31 | 2 | 0.88 | 0,93 |
| 14 | 32 | 1 | 0.94 | 0,96 |
| 15 | 32 | 1 | 0.94 | 0,96 |
| TOTAL | | | 0.90 % | 0,93 % |

* The affirmative 9 obtained a kappa considered low for the criteria adopted in this study, was removed from the instrument, for calculation.

The statements 3, 5, 11, 12, 14 and 15 showed consensus by 96% of the judges, with K of 0.94 and IVC of 0.96, standing out as the items of greater relevance of the instrument. These statements refer, respectively, to radiological protection instructions for radiodiagnostic team personnel, signage, warnings, and procedures for accidental exposures to patients, staff or the public, including radiometric survey report to staff and standards for each equipment, with harmonization of dose levels, in order to ensure the ALARA principle. Still, with respect to these affirmations, it is pointed out that the judges comments were more emphatic for affirmations 3 and 14, as follows:
There must be clear and objective work instructions, accessible to all members of the team and mandatory in their applicability. (A3)

Each equipment must have its constant and exposure table, it is also important in the case of equipment that has developed the image in a computerized way, that the calibration of the same in relation to the exposure index, to maintain especially the image quality standard. (A14)

Affirmations 1, 6, 7, 8, 10 and 13 showed consensus by 93% of the judges, with K of 0.88 and IVC of 0.93. These included the radiation protection program, area monitoring program, individual monitoring program, quality assurance program that includes maintenance program and manual of x-ray equipment, emphasizing that these should be within the reach of workers and written in Portuguese. The comments of the judges show high relevance in affirmations 1, 7 and 8, as follows, respectively:

The radiation protection program is extremely relevant for hospital practice, we rely heavily on this document, calibrated equipment, quality program and constant maintenance of analog and / or digital processors for a good examination. It only makes sense to have services that use ionizing radiation in operation, if the PR is assured for professionals and users. (A1)

The professional of radiological techniques must be co-responsible in the limitation of dose because it is he who operates the equipment that generates IR. (A7)

The method of disclosure and identification of individual radiological protection equipment should be the most appropriate to the reality of the institution where the practices are performed. Many professionals work in hospitals and also in teaching, it is important to pass this knowledge on to students. (A8)

Affirmations 2 and 4 showed consensus by 90% of the judges, with K of 0.83 and IVC of 0.90. This percentage of judges agrees that all team members should be included in the radiation protection program, as well as their assignments, responsibilities, qualifications and workload to ensure good practice of radiological protection for all. In addition, judges agree that a periodic training and refresher program of all staff on radiation protection should be conducted annually and be recorded in writing to consult the team at any time. The observations of the expert judges corroborate these statements:

Although there are periodic updates in radiodiagnosis, they are rarely about radiological protection. The training programs must be articulated with the practice, rescuing and valuing the experiences and experiences of the workers themselves. (A2)

Due to the deleterious potential of radiation when they interact with biological means, updating and raising awareness of the professionals who use these radiations in their professional activities that use these radiations in their activities is always necessary. (A4)

The affirmative 9 was the one that obtained the lowest consensus among the judges, reaching K of 0.73 and IVC of 0.83, and the recommended for this study was $k \geq 0.80$ and for this reason it was excluded. In this case, the affirmative dealt with the identification of the radiographic procedures performed, with room description, to guarantee the radiological protection of the workers exposed to the ionizing radiation of each specific place.

## Discussion:-

With the increasing demand for increasingly complex services in health services, it is evident the incorporation of new technologies and forms of work organization, especially in the radiodiagnosis and therapeutics sector. This results in a conformation of care characterized by complex treatments that can contribute to illness due to inadequate exposure to ionizing radiation, often due to lack of knowledge about radiological protection (Kim et al., 2012).

In this context, this study sought to highlight the importance of radiological protection, validating an instrument to guide practices and assist the management of radiological protection in the hospital radiodiagnosis service. The concordance percentage in the judges' evaluation was higher than the established minimum rate of 80%. This minimum value was adopted, as suggested by the literature, to verify the validity of new instruments (Wynd et al., 2003; Davies et al., 2018).

It is emphasized that to guarantee the ALARA principle the examinations must be carried out with precision, safety and quality. In addition, periodic training can help the team to adequately use available tools to improve the management of radiation protection in radiodiagnosis, in order to fully comply with the ALARA principle (Kim et al., 2012).

In this sense, insufficient knowledge about legislation and radiological protection in the training of professionals working in radiodiagnosis may be a factor that makes it difficult to involve all multiprofessional teams in the execution of what is described in the PPR. Some professionals do not even know the radiation protection program of their workplace, nor do they recognize themselves as participants in the process of implementing this document (Huhn et al., 2016).

The quality assurance program and regular audits could explore the application of radiation protection measures supported by reflections linked to evidence-based research (Hayre et al., 2018). The exposure of the worker to the different types of workloads, in the case of physical load radiodiagnosis, reflects deficits in the health work processes, whose activities are characterized by close contact with the patient, heavy and labor intensive manipulation of weights and routines, which directly implies the guarantee of service quality (de Lima Santana et al., 2016).

In this way, it can be said that within the context of evaluation of a quality service, hospital management of excellence is being performed, which guarantees not only the quality of the service, but also reinforces the importance of "Quality" in hospital management. This fact contributes to ensure the effectiveness of the care provided to the patient (Pereira et al., 2015). In the case of regular audits, the key is to involve transparency and accountability. So this does not mean that hitting is mandatory; means that there is a need to learn the best way of doing, not prioritizing a punitive conception.

Audit is not only associated with costs, it can be used to manage the quality and improvement of health work processes, facilitating the work of professionals in this area, the quality of health services and the service to users (Azevedo et al. 2018). In the case of radiodiagnosis it is possible, through audits, to visibilize and emphasize the responsibility of the professional, about the radiological protection for himself and the user.

The results show that there is a gap between theory and practice, as well as a gap in the issue of radiation protection management, since it is reported as forgotten in periodic updates of knowledge. In this perspective, in radiodiagnosis the presence of ionizing radiation is constant and the need to know the possible damages that the same cause is essential to ensure the radiological protection of professionals, patients and companions. Also, the lack of knowledge about the legislation that discusses the radiological protection plan and the unavailability of time to gather all the team collaborates for the low effectiveness of the implementation of the same in the radiodiagnosis hospital services (Huhn et al., Huhn et al. 2016b).

Thus, according to the legislation recommended, continuing education, sustained by a periodic update, should be institutionalized, since it recognizes the possible risk to which professionals working in radiodiagnosis services are exposed. Moreover, in training for the development of practical skills, it could be included improvement projects for self-care with regard to workers' health (Anderson et al., 2016; Régis et al., 2018).

The affirmative 9 was excluded from the document, for the purpose of total calculation, for not reaching $K \geq 0.80$, nevertheless the individual IVC was of 0,84, demonstrating consensus of a great part of the judges. Therefore, it is emphasized that the approach of the alternative referred to the radiographic procedures performed in each room and it is inferred that it was the question with a lower K because in the understanding of the judges this is already included in the radiation protection plan. That is, the purpose in this statement was to signal the need for a specific description of what is done in each examination room.

While many health care organizations have quality improvement departments or teams that can handle these types of efforts, it is important that organizations are familiar with processes and structures in which employees at different levels of the organization may be involved. To ensure successful outcomes of quality improvement initiatives, health professionals must work together with management, problem identification, and solution development (Kim et al., 2012).

As a limitation of the present study, it is pointed out that there were more Brazilian judges participants than Portuguese judges, although there are more Portuguese technologists. In this case, it is made explicit that there is a differentiation with respect to the professional category; in Portugal all are professionals graduated as radiology technologists, while in Brazil, there are two professional categories: the professional of medium and upper level.

## Conclusion:-
The purpose of this study was to construct and validate the contents of a hospital radiation protection management tool in two Portuguese-speaking countries, respectively, Brazil and Portugal, and at the end of the whole process, an instrument considered adequate, with consensus 90% of the judges.
Thus, it is inferred that the instrument can be used as a basic guide for the management of hospital PR and for the completion of the Descriptive Memorial for Radiological Protection in the hospital radiodiagnosis sector. In addition, it can be adapted for PR management of specific services, such as clinics that exclusively offer diagnostic imaging services.

Finally, it is considered that the construction and validation of a tool that serves to manage radiation protection contributes to the organization of diagnostic imaging services and to the effectiveness of radiological protection within them.